\documentclass[11pt,a4paper,american]{article}
\pdfoutput=1
\usepackage{lmodern}

\usepackage[T1]{fontenc}
\usepackage[latin9]{inputenc}
\usepackage{mathrsfs}
\usepackage{amsmath}
\usepackage{amssymb}
\usepackage{esint}

\makeatletter

\pdfpageheight\paperheight
\pdfpagewidth\paperwidth

\usepackage{jcappub}
\numberwithin{equation}{section}

\pdfoutput=1
\renewcommand\[{\begin{equation}}
\renewcommand\]{\end{equation}}

\makeatother

\usepackage{babel}
\begin{document}

\title{New Weyl-invariant vector-tensor theory for the cosmological constant }

\subheader{preprint number }

\author[a,b]{Pavel Jirou\v{s}ek,}

\author[a]{and Alexander Vikman}

\affiliation[a]{\em CEICO-Central European Institute for Cosmology and Fundamental Physics, }

\affiliation{\em Institute of Physics of the Czech Academy of Sciences,\\
Na Slovance 1999/2, 18221 Prague 8, Czech Republic\\
}

\affiliation[b]{\em Institute of Theoretical Physics, Faculty of Mathematics and Physics, Charles University, }

\affiliation{\em V Hole\v{s}ovi\v{c}k\'ach 2, 180 00 Prague 8, Czech Republic\\
}

\emailAdd{jirousek@fzu.cz}

\emailAdd{vikman@fzu.cz}

\abstract{We introduce a new Weyl-invariant and generally-covariant vector-tensor
theory with higher derivatives. This theory can be induced by extending
the mimetic construction to vector fields of conformal weight four.
We demonstrate that in gauge-invariant variables this novel theory
reduces to the Henneaux\textendash Teitelboim description of the unimodular
gravity. Hence, compared with the standard general relativity, our
new higher derivative vector-tensor theory has only one new global
degree of freedom - the cosmological constant. Finally we discuss
potential extensions of this vector-tensor theory. }
\maketitle

\section{Introduction }

The cosmological constant problem remains an unsolved mystery, for
reviews see e.g. \cite{Weinberg:1988cp,Martin:2012bt,Burgess:2013ara,Padilla:2015aaa}.
One of the cornerstones of this problem is a fine-tuning or (un)naturalness
of the value of the observed acceleration of our expanding universe.
However, any discussion of naturalness, fine-tuning, and especially
related anthropic reasoning \cite{Weinberg:1987dv} at least implicitly
assumes an ensemble of theories or solutions where the cosmological
constant or vacuum energy can take different values. One of the setups
where such an ensemble is realized by different solutions is the so-called
\emph{unimodular gravity} which was first proposed by Einstein almost
a century ago in \cite{Einstein:1919gv}\footnote{For more recent discussions see \cite{Weinberg:1988cp}}.
In unimodular gravity the dynamics of spacetime is given by the trace-free
part of the standard Einstein field equations. If one assumes that
matter energy-momentum tensor is conserved, the value of the cosmological
constant is given by an integration constant, for recent discussion
see e.g. \cite{Ellis:2010uc,Ellis:2013eqs}. This integration constant
is not related to the Planck and electroweak scales or any other parameters
and coupling constants of the Standard Model. This property does not
solve the cosmological constant problem, but puts it in a rather different
perspective. There are already quite a few different action principles
reproducing the dynamics of the unimodular gravity, see e.g. \cite{vanderBij:1981ym,Henneaux:1989zc,Buchmuller:1988wx,Buchmuller:1988yn,Kuchar:1991xd,Padilla:2014yea}.
The most relevant for this work is the theory by Henneaux and Teitelboim
(HT) \cite{Henneaux:1989zc}. The main advantage of this formulation
is that it is manifestly generally covariant and has a slightly simpler
formulation than in \cite{Kuchar:1991xd}. To ensure the general covariance
the HT action contains a vector field. 

On the other hand, recently another construction leading to the trace-free
equations of motion for the metric was proposed by Chamseddine and
Mukhanov in \cite{Chamseddine:2013kea} under the name \emph{Mimetic}
Gravity. This construction is dynamically equivalent to irrotational
dust minimally coupled to standard General Relativity \cite{Chamseddine:2013kea,Golovnev:2013jxa,Barvinsky:2013mea},
Hence it is more interesting for modeling dark matter. Similarly to
HT theory the energy density of this mimetic irrotational dark matter
is a Lagrange multiplier. One of the main features of this mimetic
construction is that the theory is Weyl-invariant. This Weyl-invariance
with respect to $h_{\mu\nu}=\Omega^{2}\left(x\right)h'_{\mu\nu}$
originates from the ansatz of the composite metric 
\begin{equation}
g_{\mu\nu}=h_{\mu\nu}\cdot h^{\alpha\beta}\partial_{\alpha}\phi\,\partial_{\beta}\phi\,,\label{eq:Mimetic_scalar}
\end{equation}
into the Einstein-Hilbert action. Here it is assumed that the scalar
field $\phi$ is Weyl-invariant. Different mimetic constructions with
vector fields were considered in \cite{Barvinsky:2013mea,Chaichian:2014qba,Vikman:2017gxs,Gorji:2018okn}.
All of these theories use Weyl-invariant vector fields and no one
of these constructions corresponds to the unimodular gravity. Motivated
by the HT vector-field formulation of the unimodular gravity we search
for a novel nontrivial Weyl-invariant generalization of the mimetic
ansatz for the composite metric (\ref{eq:Mimetic_scalar}) containing
a vector field $V^{\mu}$. 

\section{Mimetic vector field of conformal weight four }

In this paper we propose a new extension of the \emph{mimetic} construction
\cite{Chamseddine:2013kea} to a vector field, $V^{\alpha}$, namely
we propose to use the ansatz 

\begin{equation}
g_{\mu\nu}=h_{\mu\nu}\cdot\left(\nabla_{\alpha}^{\left.h\right)}V^{\alpha}\right)^{1/2}\text{ ,}\label{eq:mimetic_vector}
\end{equation}
where the covariant derivative, $\nabla_{\alpha}^{\left.h\right)}$,
is the Levi-Civita connection compatible with the auxiliary metric
$h_{\mu\nu}$
\[
\nabla_{\alpha}^{\left.h\right)}h_{\mu\nu}=0\text{ .}
\]
 We will call the metric $g_{\mu\nu}$ the \emph{physical metric}.
In contrast to \cite{Gorji:2018okn} the vector field $V^{\mu}$ is
not a gauge potential / connection. However, similarly to \cite{Gorji:2018okn},
with this particular form of the conformal factor in front of $h_{\mu\nu}$
the resulting theory becomes Weyl-invariant. Indeed, the Weyl transformation
of the auxiliary metric $h_{\mu\nu}$ 
\begin{equation}
h_{\mu\nu}=\Omega^{2}\left(x\right)h'_{\mu\nu}\text{ ,}\label{eq:Weyl_trans_metric}
\end{equation}
performed along with the corresponding transformation of the vector
field 
\begin{equation}
V^{\mu}=\Omega^{-4}\left(x\right)V'^{\mu}\text{ ,}\label{eq:Weyl_trans_vector}
\end{equation}
keeps the metric $g_{\mu\nu}$ invariant. This is easy to check using
\[
\nabla_{\alpha}^{\left.h\right)}V^{\alpha}=\frac{1}{\sqrt{-h}}\partial_{\alpha}\left(\sqrt{-h}V{}^{\alpha}\right)=\frac{1}{\Omega^{4}}\frac{1}{\sqrt{-h'}}\partial_{\alpha}\left(\sqrt{-h'}V'^{\alpha}\right)=\Omega^{-4}\text{ }\nabla_{\alpha}^{\left.h'\right)}V'^{\alpha}\text{ .}
\]
Unlike the constructions in \cite{Barvinsky:2013mea,Chaichian:2014qba,Vikman:2017gxs,Gorji:2018okn}
the vector field $V^{\mu}$ has \emph{conformal weight four} under
the Weyl transformations. Another crucial difference from these works
and from the original mimetic construction \cite{Chamseddine:2013kea}
is that the map (\ref{eq:mimetic_vector}) from $h_{\mu\nu}$ to $g_{\mu\nu}$
is not algebraic, but contains derivatives\footnote{This does not allow to use the inverse function theorem. The complications
due to appearance of $h_{\mu\nu,\alpha}$ are mentioned in \cite{Zumalacarregui:2013pma}. } of the auxiliary metric $h_{\mu\nu}$ as
\begin{equation}
g_{\mu\nu}=\frac{h_{\mu\nu}}{\left(-h\right)^{1/4}}\cdot\left(\partial_{\alpha}\sqrt{-h}V^{\alpha}\right)^{1/2}\,.\label{eq:non_algebraic}
\end{equation}

Substituting the ansatz (\ref{eq:mimetic_vector}) into \emph{any}
action functional $S\left[g,\Phi_{m}\right]$ (with some matter fields
$\Phi_{m}$) induces a \emph{novel} Weyl-invariant theory with the
action functional 
\begin{equation}
S\left[h,V,\Phi_{m}\right]=S\left[g\left(h,V\right),\Phi_{m}\right]\,.\label{eq:action_transform}
\end{equation}
There is also an obvious ancillary gauge invariance with respect to
\begin{equation}
V_{\mu}=V'_{\mu}+\partial_{\mu}\theta\,,\qquad\text{where}\qquad\text{\ensuremath{\Box}}\theta=0\,,\label{eq:residual_gauge}
\end{equation}
which is similar to residual gauge redundancy in the Lorenz gauge. 

Now we can plug in the ansatz (\ref{eq:mimetic_vector}) into the
Einstein-Hilbert action to obtain an action for a higher-derivative
vector-tensor theory\footnote{We use: the standard notation $\sqrt{-h}\equiv\sqrt{-\text{det}h_{\mu\nu}}$
, the signature convention $\left(+,-,-,-\right)$, and the units
$c=\hbar=1$, $M_{\text{Pl}}=\left(8\pi G_{\text{N}}\right)^{-1/2}=1$. }
\begin{equation}
S_{g}\left[h,V\right]=-\frac{1}{2}\int d^{4}x\sqrt{-h}\left[\left(\nabla_{\alpha}^{\left.h\right)}V^{\alpha}\right)^{1/2}\text{ }R\left(h\right)+\frac{3}{8}\cdot\frac{\left(\nabla_{\mu}^{\left.h\right)}\nabla_{\alpha}^{\left.h\right)}V^{\alpha}\right)^{2}}{\left(\nabla_{\sigma}^{\left.h\right)}V^{\sigma}\right)^{3/2}}\right]\text{ .}\label{eq:ShV}
\end{equation}
This is clearly a novel scalar-vector theory going beyond Horndeski
and other more recent constructions. For details see \cite{Heisenberg:2018vsk,Clifton:2011jh}.
The gravitational part of the whole theory can be more conveniently
written as 
\begin{equation}
S_{g}\left[h,V\right]=-\frac{1}{2}\int d^{4}x\sqrt{-h}\left[\sqrt{D}\text{ }R\left(h\right)+\frac{3}{8}\cdot\frac{h^{\alpha\beta}D_{,\alpha}D_{,\beta}}{D^{3/2}}\right]\text{ ,}\label{eq:ShVD}
\end{equation}
where we introduce the notation for the four-divergence 
\begin{equation}
D=\nabla_{\alpha}^{\left.h\right)}V^{\alpha}\text{ .}\label{eq:D}
\end{equation}
Under the Weyl transformations this scalar quantity has conformal
weight four 
\begin{equation}
D=\Omega^{-4}D'\text{ .}\label{eq:D_trans}
\end{equation}
It should be stressed that as a result of this procedure all matter
fields acquire a universal coupling to the vector field $V^{\alpha}$
due to the substitution (\ref{eq:action_transform}). The total action
is $S\left[h,V,\Phi_{m}\right]=S_{g}\left[h,V\right]+S_{m}\left[h,V,\Phi_{m}\right]$.

\section{Equations of motion }

Let us derive equations of motion for our novel vector-tensor theory.
\begin{equation}
\delta S=\frac{1}{2}\int d^{4}x\sqrt{-g}\left(T_{\mu\nu}-G_{\mu\nu}\right)\delta g^{\mu\nu}+\text{Boundary terms}\,,\label{eq:variation_action}
\end{equation}
where $G_{\mu\nu}$ is the Einstein tensor for $g_{\mu\nu}$ and the
energy momentum tensor of matter is defined as usual through 
\begin{equation}
T_{\mu\nu}=\frac{2}{\sqrt{-g}}\cdot\frac{\delta S_{m}}{\delta g^{\mu\nu}}\,.\label{eq:EMT_standard}
\end{equation}
The variation of the contravariant metric yields 
\begin{equation}
\delta g^{\mu\nu}=\frac{\delta h^{\mu\nu}}{\sqrt{D}}-\frac{1}{2}g^{\mu\nu}\frac{\delta D}{D}\,,\label{eq:variation_metric}
\end{equation}
where the variation of the divergence (\ref{eq:D}) can be expressed
as 
\begin{equation}
\delta D=\nabla_{\alpha}^{\left.h\right)}\delta V^{\alpha}-\frac{1}{2}h_{\alpha\beta}\,V^{\lambda}\nabla_{\lambda}^{\left.h\right)}\delta h^{\alpha\beta}\,.\label{eq:variation_divergence}
\end{equation}
Integrating by parts, neglecting the boundary terms and using $\sqrt{-g}=D\sqrt{-h}$
we obtain equation of motion for the vector field 
\begin{equation}
\frac{1}{\sqrt{-h}}\cdot\frac{\delta S}{\delta V^{\mu}}=\frac{1}{4}\text{ }\partial_{\mu}\left(T-G\right)=0\,,\label{eq:EoM_V}
\end{equation}
along with the equation of motion for the auxiliary metric
\begin{equation}
\frac{1}{\sqrt{-h}}\cdot\frac{\delta S}{\delta h^{\alpha\beta}}=\frac{\sqrt{D}}{2}\left[T_{\alpha\beta}-G_{\alpha\beta}-\frac{1}{4}g_{\alpha\beta}\left(T-G-\frac{1}{D}V^{\lambda}\partial_{\lambda}\left(T-G\right)\right)\right]=0\,,\label{eq:EoM_h}
\end{equation}
where $T=T_{\alpha\beta}g^{\alpha\beta}$ and $G=G_{\alpha\beta}g^{\alpha\beta}$.
Using the equation of motion (\ref{eq:EoM_V}) for the vector $V^{\alpha}$,
the equation of motion for the metric $h_{\mu\nu}$ transforms to
the trace-free part of the Einstein equations 
\begin{equation}
G_{\alpha\beta}-T_{\alpha\beta}-\frac{1}{4}g_{\alpha\beta}\left(G-T\right)=0\,.\label{eq:trace_free_Einstein}
\end{equation}
Crucially, both equations of motion (\ref{eq:EoM_V}) and (\ref{eq:trace_free_Einstein})
are manifestly invariant with respect to the Weyl transformations
of $h_{\mu\nu}$ and $V^{\alpha}$, as $V^{\lambda}/D=\text{inv}$
and all other quantities are expressed through manifestly gauge invariant
objects $g_{\mu\nu}$ and matter fields $\Phi_{m}$. For the later
it is convenient to consider the Weyl-invariant vector 
\begin{equation}
W^{\mu}=\frac{V^{\mu}}{\nabla_{\alpha}^{\left.h\right)}V^{\alpha}}\,.\label{eq:W}
\end{equation}
Considered as an equation on original variables equation of motion
for the vector (\ref{eq:EoM_V}) has fourth derivatives of $\left\{ h_{\mu\nu},V^{\alpha}\right\} $
while the trace-free part of the $g-$Einstein equations (\ref{eq:trace_free_Einstein})
has up to third derivatives of these original variables. 

In fact, these equations of motion are those of the so-called \emph{unimodular
}gravity. The only difference from the standard GR is that the cosmological
constant is an integration constant. Indeed, integrating the equation
of motion (\ref{eq:EoM_V}) for the vector $V^{\alpha}$ one obtains
\[
G-T=4\Lambda=\text{const}\,.
\]
Substituting this solution into the trace-free part of the Einstein
equations one derives the standard Einstein equations with the cosmological
constant $\Lambda$
\[
G_{\alpha\beta}=\Lambda g_{\alpha\beta}+T_{\alpha\beta}\,.
\]
 Hence one can say that our construction provides \emph{Mimetic Dark
Energy} or \emph{Mimetic Cosmological Constant}. 

We could guess that our mimetic theory describes unimodular gravity
by observing that in the coordinate frame\footnote{If equality holds only in a particular frame we use $"\doteq"$ instead
of $"="$. } where 
\begin{equation}
V^{\mu}\left(x\right)\doteq\frac{1}{4}\frac{x^{\mu}}{\sqrt{-h}}\,,\label{eq:special_coordinates}
\end{equation}
the determinant of the physical metric is unity, see (\ref{eq:non_algebraic})
and all quantities depend on $h_{\mu\nu}$ through 
\[
g_{\mu\nu}\doteq\frac{h_{\mu\nu}}{\left(-h\right)^{1/4}}\,.
\]
For a nice discussion on how one can construct the unimodular coordinates
where $\sqrt{-g}=1$ see \cite{vanderBij:1981ym}. 

\section{Gauge invariant variables and scalar-vector-tensor formulation }

Now we can follow a similar procedure as in \cite{Hammer:2015pcx}
and upgrade $D$ to an independent dynamical variable in order to
eliminate the second derivatives from the action, so that 
\begin{equation}
S\left[h,D,V,\lambda\right]=-\frac{1}{2}\int d^{4}x\sqrt{-h}\left[\sqrt{D}\text{ }R\left(h\right)+\frac{3}{8}\cdot\frac{h^{\alpha\beta}D_{,\alpha}D_{,\beta}}{D^{3/2}}+\lambda\left(D-\nabla_{\alpha}^{\left.h\right)}V^{\alpha}\right)\right]\text{ .}\label{eq:Constrained}
\end{equation}
Hence, we introduced a constraint with the corresponding Lagrange
multiplier and promoted theory (\ref{eq:ShV}) to a vector-tensor-scalar
theory in this way. This theory should be Weyl-invariant, as it was
the case with the original action (\ref{eq:ShV}). This requirement
forces the Lagrange multiplier, $\lambda$, to be invariant under
the Weyl transformations. In this way all matter fields acquire a
universal coupling to the scalar field $D$. 

One can further canonically normalize the kinetic term by defining
a new scalar field of conformal weight one 
\[
D=\left(\frac{\varphi^{2}}{6}\right)^{2}\,,
\]
so that the action (\ref{eq:Constrained}) takes the form 
\begin{equation}
S\left[h,\varphi,V,\lambda\right]=\int d^{4}x\sqrt{-h}\left[-\frac{1}{2}\left(\partial\varphi\right)^{2}-\frac{1}{12}\varphi^{2}\text{ }R\left(h\right)-\frac{\lambda}{72}\varphi^{4}+\frac{\lambda}{2}\cdot\nabla_{\alpha}^{\left.h\right)}V^{\alpha}\right]\,.\label{eq:canonical_normalization}
\end{equation}
The first three terms correspond to the Dirac's theory of the Weyl-invariant
gravity \cite{Dirac:1973gk}, see also \cite{Deser:1970hs}. These
terms are also the starting point for the so-called Conformal Inflation
\cite{Kallosh:2013hoa}. In our sign convention the scalar field $\varphi$
has a ghost-like kinetic term. Importantly, in contrast to \cite{Dirac:1973gk}
the would be coupling constant $\lambda$ is a Lagrange multiplier
field. All other matter fields are coupled to the physical metric
\[
g_{\mu\nu}=\frac{\varphi^{2}}{6}\cdot h_{\mu\nu}\,.
\]

The form of the action is closely related, but not identical to those
studied in \cite{Alvarez:2006uu}. The main difference is the full
diffeomorphism invariance of our action (\ref{eq:canonical_normalization})
whereas the theories studied in \cite{Alvarez:2006uu} were only invariant
with respect to transverse diffeomorphisms preserving the value of
$\sqrt{-h}$. It seems that the vector field $V^{\mu}$ (absent in
\cite{Alvarez:2006uu}) in our construction works\footnote{In another \cite{Kuchar:1991xd} generally-covariant formulation of
unimodular gravity by Kucha\v{r} instead of the vector $V^{\mu}$
there are four compensator scalar fields $X^{A}$ representing general
unimodular coordinates. Formula (\ref{eq:special_coordinates}) represents
one possible set of them and can be useful to show canonical equivalence
between our and Kucha\v{r} formulations. } as a Stückelberg, Freiherr von Breidenbach zu Breidenstein und Melsbach
field (also colloquially known as a compensator field) restoring the
full diffeomorphism invariance. However, the form of the action suggests
that the Weyl symmetry is in a sense empty (or as sometimes called
fake) in our construction and corresponds to the Noether current which
is identically vanishing, see \cite{Jackiw:2014koa,Oda:2016pok}.
We leave the clarification of this issue for a future work. 

The dynamical variables $\left\{ h_{\mu\nu},V^{\mu},\lambda,D,\right\} $
transform as 
\begin{align}
 & h_{\mu\nu}=\Omega^{2}\left(x\right)h'_{\mu\nu}\,,\label{eq:Weyl_Transform_set}\\
 & D=\Omega^{-4}\left(x\right)D'\,,\nonumber \\
 & V^{\mu}=\Omega^{-4}\left(x\right)V'^{\mu}\,,\nonumber \\
 & \lambda=\lambda'\text{ .}\nonumber 
\end{align}
Instead of these dynamical variables one can introduce a new set of
independent dynamical variables $\left\{ g_{\mu\nu},W^{\mu},\Lambda,D\right\} $,
where the first three 
\begin{align}
 & g_{\mu\nu}=D^{1/2}\,h{}_{\mu\nu}\,,\label{eq:gauge_invariant_variables}\\
 & W^{\mu}=D^{-1}\,V^{\mu}\,,\nonumber \\
 & \Lambda=\frac{\lambda}{2}\text{ ,}\nonumber 
\end{align}
are gauge invariant. This field-redefinitions resemble the Weyl transformations
with $\Omega=D^{1/4}$, except we do not transform $D$ and consequently
do not reduce the dimensionality of the phase space. Hence this transformation
is different from fixing the gauge where $D=1$, even though the variables
$W^{\mu}$ and $g_{\mu\nu}$ are equal to the corresponding variables
in this gauge. 

In this way the divergence transforms 
\[
\nabla_{\alpha}^{\left.h\right)}V^{\alpha}=\frac{D}{\sqrt{-g}}\partial_{\alpha}\left(\sqrt{-g}\,W^{\alpha}\right)=D\,\nabla_{\mu}^{\left.g\right)}W^{\mu}\text{ .}
\]
Performing this field redefinition in (\ref{eq:Constrained}) one
obtains 
\begin{equation}
S\left[g,W,\Phi_{m}\right]=\int d^{4}x\sqrt{-g}\left[-\frac{1}{2}R\left(g\right)+\Lambda\left(\nabla_{\mu}^{\left.g\right)}W^{\mu}-1\right)\right]+S_{m}\left[g,\Phi_{m}\right]\text{ .}\label{eq:H-T_action}
\end{equation}
This action functional does not depend anymore on the scalar field
$D$, but only on gauge invariant dynamical variables (\ref{eq:gauge_invariant_variables}).
In fact, as a result of this transformation we obtained the Henneaux-Teitelboim
representation \cite{Henneaux:1989zc} of the unimodular gravity.
Indeed, the variation of this action with respect to the vector field
$W^{\mu}$ implies that $\Lambda$ is a constant of integration (global
degree of freedom):
\[
\frac{1}{\sqrt{-g}}\cdot\frac{\delta S}{\delta W^{\mu}}=-\partial_{\mu}\Lambda=0\text{ ,}
\]
while the variation with respect to the metric gives the Einstein
equations with the cosmological constant $\Lambda$ 
\[
\frac{2}{\sqrt{-g}}\cdot\frac{\delta S}{\delta g^{\mu\nu}}=T_{\mu\nu}+\Lambda g_{\mu\nu}-G_{\mu\nu}=0\text{ .}
\]
Finally there is a constraint 
\begin{equation}
\frac{1}{\sqrt{-g}}\cdot\frac{\delta S}{\delta\Lambda}=\nabla_{\mu}^{\left.g\right)}W^{\mu}-1=0\,.\label{eq:current_nonconservation}
\end{equation}
The constraint equation (\ref{eq:current_nonconservation}) per construction
becomes identity in terms of original fields $\left\{ h_{\mu\nu},V^{\alpha}\right\} $
and does not provide any new information regarding the dynamics. In
electrodynamics one faces a similar situation with $\nabla_{\mu}^{\left.g\right)}F^{\mu\nu}$
which is identically conserved per construction. 

The constraint equation (\ref{eq:current_nonconservation}) can be
considered as a non-conservation of the current $W^{\mu}$. This equation
only allows to find the evolution of the corresponding charge - the
global mode defined on a foliation of the spacetime as 
\begin{equation}
\mathscr{T}\left(t\right)=\int d^{3}\mathbf{x}\sqrt{-g}\,W^{t}\left(t,\mathbf{x}\right)\,,\label{eq:global_time}
\end{equation}
which is often called ``cosmic time'' or four-dimensional spacetime
volume, see \cite{Henneaux:1989zc}. Indeed, using (\ref{eq:current_nonconservation})
we can calculate 
\begin{align*}
 & \dot{\mathscr{T}}\left(t\right)=\int d^{3}\mathbf{x}\partial_{t}\left(\sqrt{-g}\,W^{t}\left(t,\mathbf{x}\right)\right)=\int d^{3}\mathbf{x}\left(\sqrt{-g}-\partial_{i}\left(\sqrt{-g}W^{i}\right)\right)=\\
 & =\int d^{3}\mathbf{x}\sqrt{-g}-\oint_{\mathcal{B}}ds_{i}\sqrt{-g}W^{i}\,,
\end{align*}
where the last integral is taken over the boundary surface $\mathcal{B}$
of the three-dimensional space. If there is no flux of $W^{i}$ through
the boundary surface, then 
\[
\mathscr{T}\left(t_{2}\right)-\mathscr{T}\left(t_{1}\right)=\int_{t_{1}}^{t_{2}}dt\int d^{3}\mathbf{x}\sqrt{-g}\,.
\]
It is worth noting that one can write the ''cosmic time'' $\mathscr{T}\left(t\right)$
in terms of $\left\{ h_{\mu\nu},V^{\alpha}\right\} $:
\begin{equation}
\mathscr{T}\left(t\right)=\int d^{3}\mathbf{x}\sqrt{-h}\,V^{t}\left(t,\mathbf{x}\right)\,,\label{eq:global_degree}
\end{equation}
as the tensor density $\sqrt{-h}V^{\mu}$ is invariant under the Weyl
transformations (\ref{eq:Weyl_Transform_set}) and remains invariant
under the field redefinition (\ref{eq:gauge_invariant_variables}).
In the special coordinate system (\ref{eq:special_coordinates}) the
charge expression takes a particularly simple form 
\[
\mathscr{T}\left(t\right)\doteq t\int d^{3}\mathbf{x}\,.
\]

Clearly there is still a lot of gauge redundancy in the action (\ref{eq:H-T_action}),
as it does not allow to find all components of $W^{\mu}$, but only
the global mode $\mathscr{T}\left(t\right)$. To find the evolution
of the global mode one has to specify conditions for normal components
of $W^{i}$ to the spatial boundary surface $\mathcal{B}$ at all
times and initial $W^{t}\left(t_{1},\mathbf{x}\right)$ (or final
$W^{t}\left(t_{2},\mathbf{x}\right)$) charge density. Of course one
can specify both, the initial $W^{t}\left(t_{1},\mathbf{x}\right)$
and the final $W^{t}\left(t_{2},\mathbf{x}\right)$, though in that
case the boundary conditions for $W^{i}$ should be chosen consistently
so that the flux of the current $W^{i}$ could compensate for the
changes in the charge additional to the the four-volume of the spacetime
between two Cauchy hypersurfaces: 
\[
\mathscr{T}\left(t_{2}\right)-\mathscr{T}\left(t_{1}\right)=\int_{t_{1}}^{t_{2}}dt\int d^{3}\mathbf{x}\sqrt{-g}-\int_{t_{1}}^{t_{2}}dt\oint_{\mathcal{B}}ds_{i}\sqrt{-g}W^{i}\,.
\]
Of course very different charge densities $W^{t}\left(t,\mathbf{x}\right)$
can still correspond to the same global charge $\mathbf{\mathscr{T}}\left(t\right)$. 

\section{Conclusion and Discussion}

We proposed a new vector-tensor theory (\ref{eq:ShV}) with a vector
field $V^{\mu}$ of conformal weight four:
\begin{equation}
S_{g}\left[h,V\right]=-\frac{1}{2}\int d^{4}x\sqrt{-h}\left[\left(\nabla_{\alpha}^{\left.h\right)}V^{\alpha}\right)^{1/2}\text{ }R\left(h\right)+\frac{3}{8}\cdot\frac{\left(\nabla_{\mu}^{\left.h\right)}\nabla_{\alpha}^{\left.h\right)}V^{\alpha}\right)^{2}}{\left(\nabla_{\sigma}^{\left.h\right)}V^{\sigma}\right)^{3/2}}\right]\text{ .}\label{eq:our_theory_conclusion}
\end{equation}
 Notably the Weyl-symmetry is considered to be a desirable and intriguing
property in physics, see e.g. \cite{Lucat:2016eze}. This higher-derivative
Weyl-invariant theory is highly degenerate and has only one global
degree of freedom (\ref{eq:global_degree})
\begin{equation}
\mathscr{T}\left(t\right)=\int d^{3}\mathbf{x}\sqrt{-h}\,V^{t}\left(t,\mathbf{x}\right)\,,\label{eq:global_dof_concluison}
\end{equation}
whose canonical momentum is the cosmological constant $\Lambda$.
This global degree of freedom is Weyl-invariant. We obtained this
theory by making a mimetic substitution 

\begin{equation}
g_{\mu\nu}=h_{\mu\nu}\cdot\left(\nabla_{\alpha}^{\left.h\right)}V^{\alpha}\right)^{1/2}\text{ ,}\label{eq:ansatz_metric_conclusion}
\end{equation}
into the Einstein-Hilbert action. 

Further we reformulated this theory as a Weyl-invariant scalar-vector-tensor
gravity (\ref{eq:canonical_normalization}), which closely resembles
the Dirac's theory of the Weyl-invariant gravity \cite{Dirac:1973gk}.
However, our formulation has an additional constraint and a vector
field of different conformal weight. Then we introduced gauge-invariant
local variables (\ref{eq:gauge_invariant_variables}) and found that
our theory reduces to the generally covariant Henneaux-Teitelboim
representation \cite{Henneaux:1989zc} of \emph{unimodular} gravity.
In contrast to other formulations of unimodular gravity our action
(\ref{eq:ShV}) has manifest i) Weyl-invariance and ii) general covariance,
while there are iii) no explicit constraints imposed using Lagrange
multipliers. The price for the combination of all these three properties
is the presence of higher derivatives in the action. Despite of these
higher derivatives the theory does not suffer from the Ostrogradsky
ghosts \cite{Ostrogradsky:1850fid} in the standard sense. Indeed,
as it is in the standard Ostrogradsky prescription, the Henneaux-Teitelboim
theory is linear in the canonical momentum $\Lambda$, but for each
solution the momentum stays constant in the whole spacetime. 

Vector-tensor theories are quite popular in the context of modeling
dark energy and dark matter phenomena, for recent reviews see e.g.
\cite{Heisenberg:2018vsk,Clifton:2011jh}. Clearly our vector field
is not a U(1) gauge potential. Our theory goes beyond Horndeski's
most general construction for the U(1) vector fields with second order
equations of motion \cite{Horndeski:1976gi}. Neither can one find
our theory in more general p-form constructions \cite{Deffayet:2010zh,Deffayet:2016von,Deffayet:2017eqq}.
Moreover our construction goes beyond popular generalized Proca vector-tensor
theories \cite{Heisenberg:2014rta} and Einstein aether models \cite{Jacobson:2000xp}
where the U(1) invariance is broken, and goes even beyond further
extended vector-tensor theories \cite{Kimura:2016rzw,Heisenberg:2016eld}.
Also our construction is principally different from other mimetic
vector models \cite{Barvinsky:2013mea,Chaichian:2014qba,Vikman:2017gxs,Gorji:2018okn}.
A crucial difference is that our contravariant vector field has conformal
weight four contrary to the ordinary Weyl-invariant covariant vector
fields of weight zero used in the previous constructions. Another
difference is that we have derivatives of the metric inside of the
mimetic transformation (\ref{eq:ansatz_metric_conclusion}). 

Another interesting feature of our formulation of the unimodular gravity,
which is common with \cite{Henneaux:1989zc,Kuchar:1991xd}, is a spontaneous
breaking of the Lorentz symmetry. Indeed, in our construction (\ref{eq:ansatz_metric_conclusion})
vanishing $V^{\mu}$ corresponds to a singularity. In Weyl-invariant
or HT formulation the persistent presence of $W^{\mu}$ is enforced
by the constraint (\ref{eq:current_nonconservation}). Clearly this
Lorentz-symmetry breaking is not relevant as far as it does not propagate
to the Standard Model fields. Interestingly, one can reproduce the
dynamics of the original Chamseddine-Mukhanov scalar mimetic dark
matter via Lorentz-symmetry breaking in a so-called pre-geometric
setup \cite{Zlosnik:2018qvg}, where the spacetime manifold appears
only via Lorentz-symmetry breaking. One can wonder whether a different
pre-geometric setup can provide our \emph{Mimetic Cosmological Constant}
or maybe evolving\emph{ Mimetic Dark Energy} along with their initial
data. 

Different formulations of the same classical theory can correspond
to distinct quantum theories. For unimodular gravity a relevant discussion
on this point can be found in e.g. \cite{Fiol:2008vk,Padilla:2014yea}.
These differences can be important for quantum vacuum energy and for
the UV structure of the theory. Moreover, formulations of the same
theory in terms of different dynamical variables are relevant for
potential modifications and extensions. In particular, these modifications
are interesting in any attempt to dynamically compensate the cosmological
constant. On the other hand, extensions can model deviations from
an exact cosmological constant to novel forms of evolving vacuum energy.
Hence suggesting another formulation of the unimodular gravity can
be useful also in this regard. 

Finally we would like to mention further ways of generalizing our
setup. Scalar field mimetic models can be extended by plugging in
the mimetic ansatz into actions already containing different scalar-field
operators. In particular, this procedure yields phenomenologically
interesting theories for the operators $V\left(\phi\right)$ (see
e.g. \cite{Chamseddine:2014vna,Lim:2010yk}) and $\gamma\left(\phi\right)\left(\Box\phi\right)^{2}$,
see e.g. \cite{Chamseddine:2014vna,Capela:2014xta,Mirzagholi:2014ifa,Ramazanov:2015pha,Casalino:2018wnc}.
The latter operators are rather constrained phenomenologically \cite{Babichev:2016jzg,Casalino:2018wnc,Ramazanov:2016xhp}
especially as they can introduce mild ghost instabilities, see \cite{Ramazanov:2016xhp}
and \cite{Chaichian:2014qba,Ijjas:2016pad,Takahashi:2017pje,Langlois:2018jdg}.
Different extensions are also interesting, as they can point out directions
to embed or UV complete the theory. To extend our vector-tensor theory,
one can start from any progenitor theory with dynamical variables
$\{g_{\mu\nu},W^{\mu}\}$ and some matter fields $\Phi_{m}$ and perform
\emph{simultaneous }transformation of metric (\ref{eq:ansatz_metric_conclusion})
and of the vector field (\ref{eq:W}): 
\begin{equation}
W^{\mu}=\frac{V^{\mu}}{\nabla_{\alpha}^{\left.h\right)}V^{\alpha}}\,.\label{eq:composite_vector_conclusions}
\end{equation}
After substituting these transformed composite objects into the progenitor
tensor (or vector-tensor or even scalar-vector-tensor) theory we obtain
a new vector-tensor theory with the action
\[
S\left[h,V,\Phi_{m}\right]=S\left[g\left(h,V\right),W\left(h,V\right),\Phi_{m}\right]\,.
\]
This induced novel vector-tensor theory (with some external matter
fields $\Phi_{m}$) is Weyl-invariant per construction. After transition
back to the Weyl-invariant variables (\ref{eq:gauge_invariant_variables})
one just adds a constraint term (\ref{eq:current_nonconservation}),
$\Lambda\left(\nabla_{\mu}^{\left.g\right)}W^{\mu}-1\right)$, to
the original action $S\left[g,W,\Phi_{m}\right]$. For example, extending
our model by adding the standard kinetic term $-\frac{\text{1}}{4}F_{\mu\nu}F^{\mu\nu}$
with the usual field tensor $F_{\mu\nu}=\partial_{\mu}W_{\nu}-\partial_{\nu}W_{\mu}$
to the progenitor Einstein-Hilbert action and making the combined
mimetic ansatz (\ref{eq:ansatz_metric_conclusion}),(\ref{eq:composite_vector_conclusions})
generates a new Weyl\textendash invariant gauge theory preserving
even the residual gauge symmetry (\ref{eq:residual_gauge}). One can
also use a different (e.g. with curvature corrections) progenitor
gravitational Lagrangian instead of the standard Einstein-Hilbert
one. 

One can further expand the story by using various tensor fields of
other conformal weights. For each progenitor field $\Psi$ which we
want to transform to a field of conformal weight $k$ one should make
a substitution of a composite field with $\Psi=\psi\left(\nabla_{\alpha}^{\left.h\right)}V^{\alpha}\right)^{-k/4}$. 

Another interesting generalization is an extension of the conformal
mimetic ansatz (\ref{eq:ansatz_metric_conclusion}) to more general
\emph{disformal} transformations \cite{Bekenstein:1992pj} where instead
of the usual $\partial_{\mu}\phi$ one exploits the vector field $V^{\mu}$.
For instance the transformation 
\[
g_{\mu\nu}=h_{\mu\nu}\cdot\left(\nabla_{\alpha}^{\left.h\right)}V^{\alpha}\right)^{1/2}+\left(\nabla_{\alpha}^{\left.h\right)}V^{\alpha}\right)^{-1}\,V_{\mu}\,V_{\nu}\,,
\]
 still generates new Weyl-invariant theories. Whether the induced
theories can describe interesting physics, similarly to the setup
presented in the paper, remains to be seen and is an interesting open
question. \newpage{}

\acknowledgments It is a pleasure to thank Ippocratis Saltas for
useful discussions, Atsushi Naruko for a helpful correspondence and
Tom Z\l{}o\'snik for valuable comments on the first draft of the
paper. A.V. acknowledges support from the J. E. Purkyn\v{e} Fellowship
of the Czech Academy of Sciences. The work of P. J. and A.V. was supported
by the funds from the European Regional Development Fund and the Czech
Ministry of Education, Youth and Sports (M\v{S}MT): Project CoGraDS
- CZ.02.1.01/0.0/0.0/15\_003/0000437. \\

\bibliographystyle{utphys}
\addcontentsline{toc}{section}{\refname}\bibliography{Unimod}

\end{document}